\documentclass[twocolumn,superscriptaddress,floatfix,nobalancelastpage,10pt,pra]{revtex4-2}

\usepackage{graphicx}
\usepackage{dcolumn}
\usepackage{bm}
\usepackage{color}
\usepackage{soul}
\usepackage{amsmath}
\usepackage[normalem]{ulem}
\usepackage{tabularx}
\usepackage{comment}
\usepackage{ragged2e}
\usepackage[caption=false]{subfig}
\usepackage{placeins}
\usepackage{silence}

\setstcolor{red}

\newcommand{\equalcontrib}{These authors contributed equally to this work.}

\begin{document}

\WarningFilter{LaTeX}{A float is stuck}
\WarningFilter{revtex4}{Deferred float stuck during \string\clearpage\space processing.}

	\preprint{APS/123-QED}
	
	\title{Verification of continuous variable entanglement with undetected photons
	}
	
	\author{Sanjukta Kundu}
    \thanks{\equalcontrib}
    \affiliation{Institute of Experimental Physics, University of Warsaw. Pasterura 5 Warsaw}
   
	\author{Balakrishnan Viswanathan}
     \thanks{\equalcontrib}
    \affiliation{Department of Physics, Oklahoma State University, 145 Physical Sciences Bldg, Stillwater, Oklahoma 74078}
    \affiliation{Current address: Optics and Quantum Information Group, The Institute of Mathematical Sciences,\\ CIT Campus, Chennai 600113, India}

	\author{Pawel Szczypkowski}
	\affiliation{Institute of Experimental Physics, University of Warsaw. Pasterura 5 Warsaw}
	
	\author{Gabriela Barreto Lemos}
	\affiliation{Instituto de Fisica, Universidade Federal do Rio de Janeiro, Av. Athos da Silveira Ramos, Rio de Janeiro, CP: 68528, Brazil}
	
	\author{Mayukh Lahiri}
	\email{mlahiri@okstate.edu}
	\affiliation{Department of Physics, Oklahoma State University, 145 Physical Sciences Bldg, Stillwater, Oklahoma 74078}

    \author{Radek Lapkiewicz} 
    \email{radek.lapkiewicz@fuw.edu.pl}
	\affiliation{Institute of Experimental Physics, University of Warsaw. Pasterura 5 Warsaw}

	\date{\today}
	
	\begin{abstract}
    We verify transverse spatial entanglement of photon-pairs generated in spontaneous parametric down conversion using a nonlinear interferometric technique without relying on any coincidence detection. We experimentally demonstrate the violation of the Einstein-Podolsky-Rosen criterion and of the Mancini-Giovannetti-Vitali-Tombesi criterion using single photon interference of one of the photons of the pairs. We also provide a comprehensive theoretical analysis. The experimental results that we have obtained show good agreement with the theoretical values. Our method performs well under experimental losses and can be applied to highly non-degenerate sources, where there are no suitable detectors for one of the photons in the quantum state and our method could also be extended to the discrete degrees of freedom to certify high-dimensional (OAM) entanglement. \\[2 pt]
    
    \end{abstract}
	
	\maketitle
	
	\noindent

\section{Introduction}
Entanglement lies at the foundation of quantum mechanics and serves as a central resource across modern quantum technologies, enabling advances in imaging, communication, computation, and metrology. In continuous-variable photonic systems, spatial entanglement in position and momentum (see \cite{walborn2010spatial} and references therein) provides access to high-dimensional Hilbert spaces and enhanced information-encoding capabilities. These  transverse spatial correlations have opened opportunities across various quantum technologies, including quantum imaging \cite{defienne2024}, quantum metrology \cite{triggiani2024estimation, triggiani2025momentum}, and quantum key distribution \cite{almeida2005experimental}. However, conventional methods for verifying such correlations rely on coincidence detection of both photons in a pair \cite{howell2004realization, aspden2013epr,moreau2014einstein, Hor-meyll}, which can be experimentally challenging, particularly for highly non-degenerate sources where efficient detectors may not exist for one of the photons. This limitation motivates us the development of alternative strategies for certifying continuous-variable entanglement that do not depend on coincidence measurements and remain robust under realistic experimental imperfections.
\par
In contrast to the traditional approach of characterizing entanglement in any two-photon system, which relies on detection of both photons via coincidence measurement (Fig.~\ref{fig:principle}a) \cite{freedman1972experimental,aspect1982experimental,ou1992realization,James2001,giustina-2015-significant}, we 
take an interferometric approach (Fig.~\ref{fig:principle}b) that dispenses with the use of coincidence detection. Our method relies on a phenomenon often called \emph{induced coherence without induced emission} \cite{zou1991induced,WZM-ind-coh-PRA}. Techniques based on this phenomenon have recently been applied to measure entanglement in two-qubit photonic systems without detecting one of the photons \cite{lahiri2021characterizing,lemos2023one,rajeev2023single,rajeev2025complete}. Here, we demonstrate both theoretically and experimentally that it is possible to characterize spatial (momentum-position) entanglement in a bipartite photonic system using only single-photon counting rate without any postselection. 
\par
Our method provides practical advantage over traditional techniques when one of the two photons lies in a spectral range for which adequate detectors are not available. Moreover, it performs well even under high experimental losses. The method can also be implemented without using single-photon detectors, i.e., using standard cameras, such as CCD or sCMOS.
\par
The paper is organized as follows: In Sec.~\ref{Sec:principle}, we outline the principle of our method. In Sec.~\ref{Sec:experiment}, we describe the experiment in detail followed by a rigorous discussion of the underlying theory in Sec.~\ref{Sec:Theory}. The results are presented in Sec.~\ref{Sec:Results} and Sec.~\ref{Sec:discuss} provides a brief discussion of our method. We finally summarize our work in Sec.~\ref{Sec:Summary}.
\vspace{-0.6em}
\section{Principle}\label{Sec:principle}    
    
Let us consider a photon pair that can be correlated in momentum and position. The two photons constituting a pair will be called signal ($S$) and idler ($I$). We represent the transverse momenta of signal and idler photons at the source plane by $\hbar\textbf{k}_S$ and $\hbar\textbf{k}_I$, respectively, where $\textbf{k}_S\equiv (k_{Sx},k_{Sy})$ and $\textbf{k}_I\equiv (k_{Ix},k_{Iy})$. Likewise, the transverse positions of signal and idler photons at the source are denoted by $\boldsymbol{\rho}_{S}\equiv (x_S,y_S)$ and $\boldsymbol{\rho}_I\equiv (x_I,y_I)$, respectively. The photon pair is entangled in position-momentum if the following Einstein-Podolsky-Rosen (EPR) type criterion is violated \cite{reid1989demonstration,howell2004realization, reid2009colloquium,edgar2012imaging} 
	 \begin{align} \label{EPR def}
	\Delta^{2}(\boldsymbol{\rho}_{I} | \boldsymbol{\rho}_{S}) \, \Delta^{2}(\textbf{k}_{I} | \textbf{k}_{S}) &> \frac{1}{4},
	\end{align}
where $\Delta^{2}(\xi_I|\xi_S)$ is the \emph{conditional variance}, representing uncertainty in the measurement of variable $\xi_I$, conditional on the measurement of variable $\xi_S$. 
\begin{figure}[htp]
		\centering 
		\includegraphics[width=.98\linewidth]{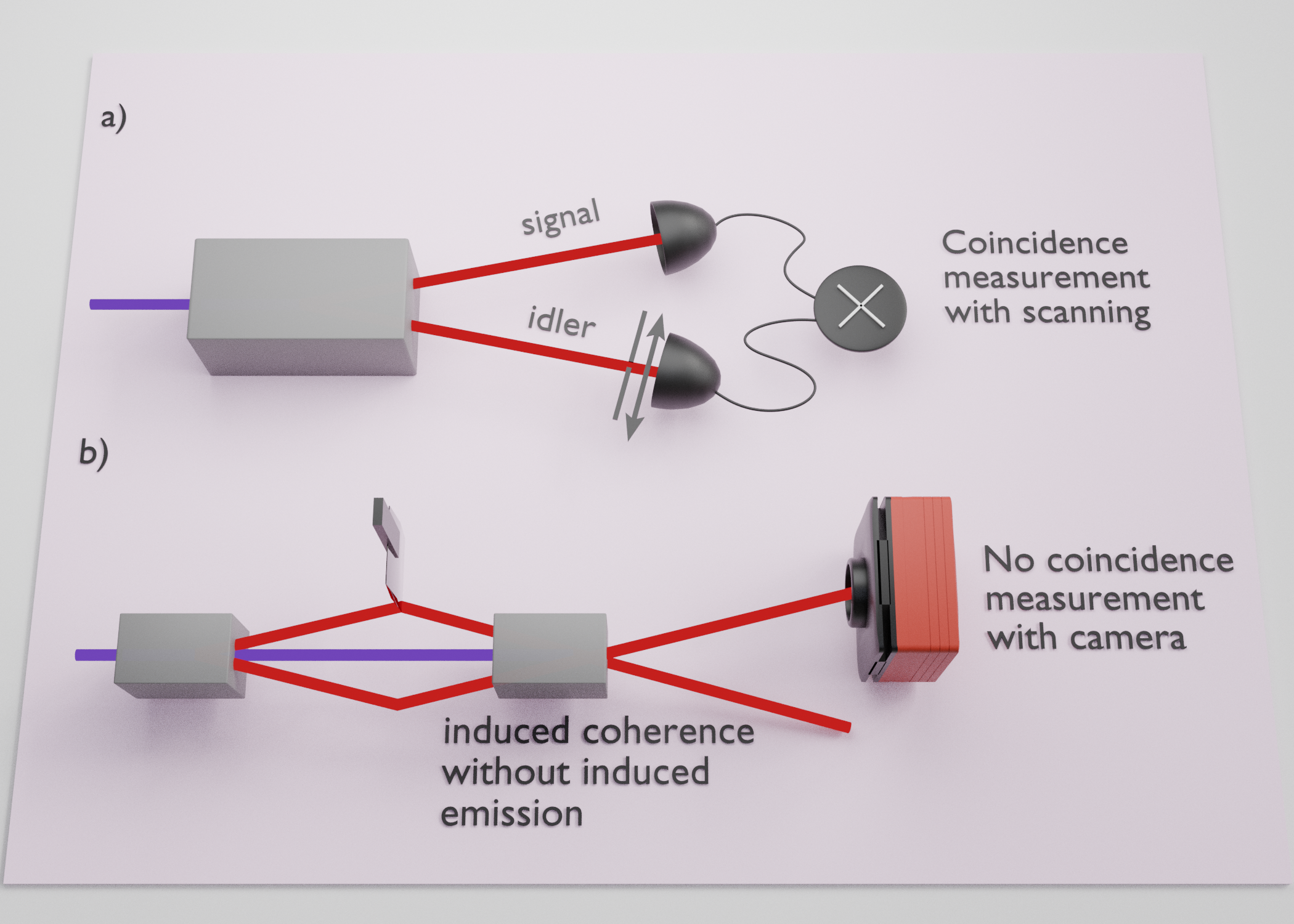}
		\caption{Comparison between the principle of standard coincidence-based entanglement measurement and our method using induced coherence.
(a) Standard approach: Spatial entanglement of photon pairs generated via SPDC is typically measured by detecting both signal and idler photons in coincidence. Scanning detectors are usually used to record joint probability distributions in position or momentum bases.
(b) Our approach: Using an induced coherence interferometer, only the signal photon is detected on an EMCCD camera, while the idler photon remains undetected. By analyzing the interferometric image of the knife-edge in position and momentum space, we retrieve the same conditional variances without coincidence detection.}
		\label{fig:principle}
	\end{figure}

\par
Measuring $\Delta^{2}(\boldsymbol{\rho}_{I} | \boldsymbol{\rho}_{S})$ and $\Delta^{2}(\textbf{k}_{I} | \textbf{k}_{S})$ is the cornerstone of testing this criterion. We realize these measurements by analyzing the single-photon interference pattern generated by \textit{induced coherence without induced emission} \cite{zou1991induced,WZM-ind-coh-PRA}. 
We now consider two mutually coherent twin-photon sources, each of which can generate the target quantum state. These can be at two spatially separated locations, for example, two nonlinear crystals pumped by two mutually coherent laser beams (Fig.~\ref{fig:principle}b). These two sources can also be at the same location \cite{herzog1994frustrated}, for example, in a single nonlinear crystal pumped from two directions by two mutually coherent laser beams (Figs.~\ref{fig_schematic}a and \ref{fig_schematic}b). For simplicity of discussion, we choose the former arrangement to describe the principle. For compactness and the stability of the interferometer, the latter approach is taken in the experiment. The same theoretical treatment is applicable to both scenarios. 
\par
The paths of the photon pair generated by individual sources are perfectly aligned (Fig.~\ref{fig:principle}b), such that it is impossible to determine the ``which-way information'' corresponding to a detected photon \cite{zou1991induced,WZM-ind-coh-PRA,herzog1994frustrated}, resulting in a single-photon interference pattern. When we place an object in the path of the undetected photon between the two sources, the spatial information of the object appears in the interference pattern \cite{lemos2014quantum,lahiri2015theory,viswanathan2021position,kviatkovsky2022mid}. We show that if the object is known, one can determine the conditional variances, $\Delta^{2}(\boldsymbol{\rho}_{I} | \boldsymbol{\rho}_{S})$ and $\Delta^{2}(\textbf{k}_{I} | \textbf{k}_{S})$, from the single-photon interference data. Furthermore, the same data allow us to test the violation of the Mancini-Giovannetti-Vitali-Tombesi (MGVT) criterion \cite{MGVT2002PRL}, whose relevant form in our case is
    \begin{equation} \label{MGVT criteria-main}
	 \Delta^{2}\textbf{K}_{+} \, \Delta^{2}\pmb{\rho}_{-} \geq 1, 
	\end{equation}
where $\Delta^{2} \textbf{K}_{+}\equiv (\Delta^{2} K_{x+},\Delta^{2} K_{y+})$ and $\Delta^{2} \pmb{\rho}_{-}\equiv (\Delta^{2} X_{-},\Delta^{2} Y_{-})$ are the variances of $\textbf{K}_{+} = \textbf{k}_{I}+\textbf{k}_{S}$ and $\pmb{\rho}_{-} = \boldsymbol{\rho}_{S} - \boldsymbol{\rho}_{I}$, respectively.

\section{Description of the Experiment}\label{Sec:experiment}

\begin{figure*}[t]
    \centering
    \includegraphics[width=1.0\linewidth]{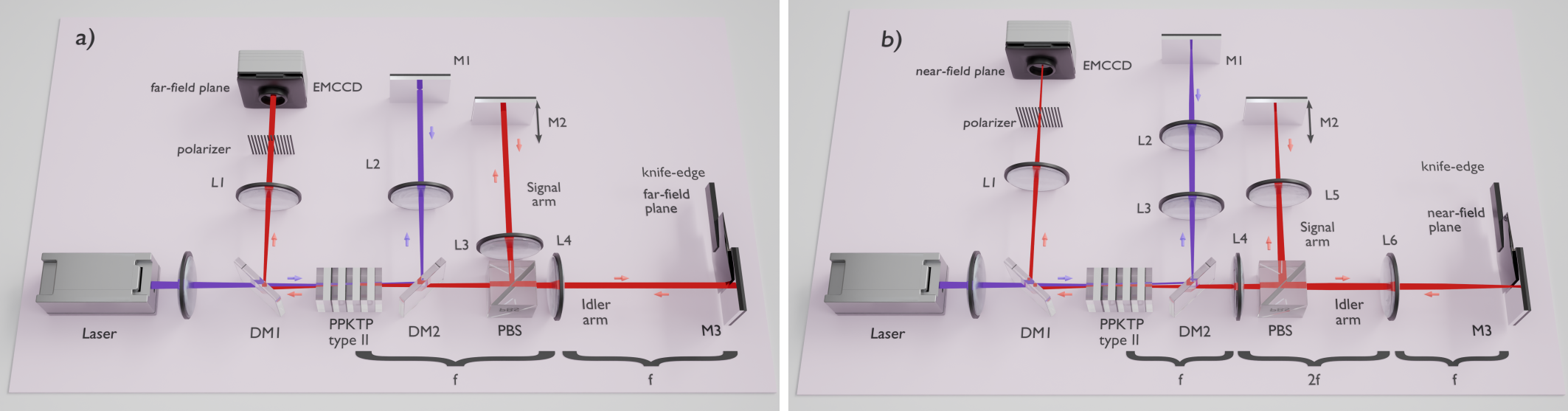}
    \caption{\justifying
    Experimental setup for measuring the edge-spread function (ESF) in (a) momentum space and (b) position space. A 405~nm pump beam generates degenerate, orthogonally polarized photon pairs via type-II SPDC in a ppKTP crystal. Interference of the generated photon is enabled by induced coherence. A knife edge is placed in the idler arm to perform edge measurements in either the far-field (a) or near-field (b) of the crystal. \newline (a) Momentum-space configuration: All lenses used have 200~mm focal lengths. The pump is focused onto the crystal, generating SPDC photon pairs. DM2 separates the forward-propagating pump from the signal and idler photons. Lens L2 projects the crystal’s far-field onto mirror M1 and simultaneously focuses the back-propagating pump (reflected from M1) into the crystal. After the PBS, lenses L3 (signal arm) and L4 (idler arm) project the far-field onto mirrors M2 and M3, respectively. The knife edge is placed just before M3 in the idler arm, in the far-field plane of the crystal. The back-propagating signal and idler are separated from the pump by DM1, and lens L1 images the crystal’s far-field onto the EMCCD camera. A polarizer selects the signal photon polarization before detection.\newline (b) Position-space configuration: All lenses have 100~mm focal lengths. The pump is focused onto the crystal, where SPDC photon pairs are generated. DM2 separates the forward-propagating pump from the down-converted photons. Lenses L3 and L4 image the crystal’s near-field onto M1 and simultaneously focus the back-propagating pump (from M1) into the crystal. Additional 4f imaging systems (L4–L5 for the signal arm and L4–L6 for the idler arm) project the near-field onto mirrors M2 and M3. The signal and idler photons are separated by a PBS placed after L4. The knife edge is positioned just before M3 in the near-field plane of the crystal. The back-propagating signal and idler are separated from the pump at DM1, and a 2f–2f imaging system using L1 maps the crystal’s near-field onto the EMCCD camera. A polarizer selects the detected signal photon.
    }
    \label{fig_schematic}
\end{figure*}
    
 Toward certifying the position-momentum entanglement experimentally with our method, we build a double-pass non-linear interferometer using the collinear type-II spontaneous parametric down-conversion (SPDC) process (Figs.~\ref{fig_schematic}a and \ref{fig_schematic}b). Our interferometer is based on the frustrated two-photon creation interferometer \cite{herzog1994frustrated}. We pump a 2$~$mm long ($L$) periodically poled potassium titanyl phosphate (ppKTP) nonlinear crystal maintained at $32^\circ\text{C}$ with $405~nm$ ($\lambda_p$) laser. The pump is focused on the crystal with a waist of $108~\mu m$ ($w_0$) and it  results in the creation of degenerate, orthogonally polarized twin-photon pairs (signal and idler) at a wavelength $810~\text{nm}$ into paths $S_1$ and $I_1$. We separate the pump beam from the twin-photon pair with a dichroic mirror. Sequentially, we employ a polarizing beam splitter to separate the signal from the idler. The pump, signal, and idler are directed along three distinct paths and are reflected back by mirrors $M_1$, $M_2$, and $M_3$, respectively, placed at the end of the three interferometric arms. After getting reflected back from the mirrors, both signal and idler photons and the pump pass through the crystal. The back-propagating pump beam can generate the second set of twin-photon pairs (signal and idler) into paths $S_2$ and $I_2$. Thus, the two-possible ways (forward and backward) of creating photon pairs can interfere. It is important to ensure that the photon paths ($S_1$, $I_1$) corresponding to the first pass of the pump beam are aligned precisely, and overlap with the photon paths ($S_2$, $I_2$) corresponding to the second pump pass. We use a short-pass dichroic mirror to separate the signal and idler photons from the pump beam and direct them toward an EMCCD camera. Once we assure that the alternative paths overlap and that the photon pairs, emitted as the pump passes through the crystal twice, are indistinguishable, we observe the interference of either the signal or idler photons. 
\par
\begin{figure*}[!htp]
		\centering 
		\includegraphics[width=1.0\linewidth]{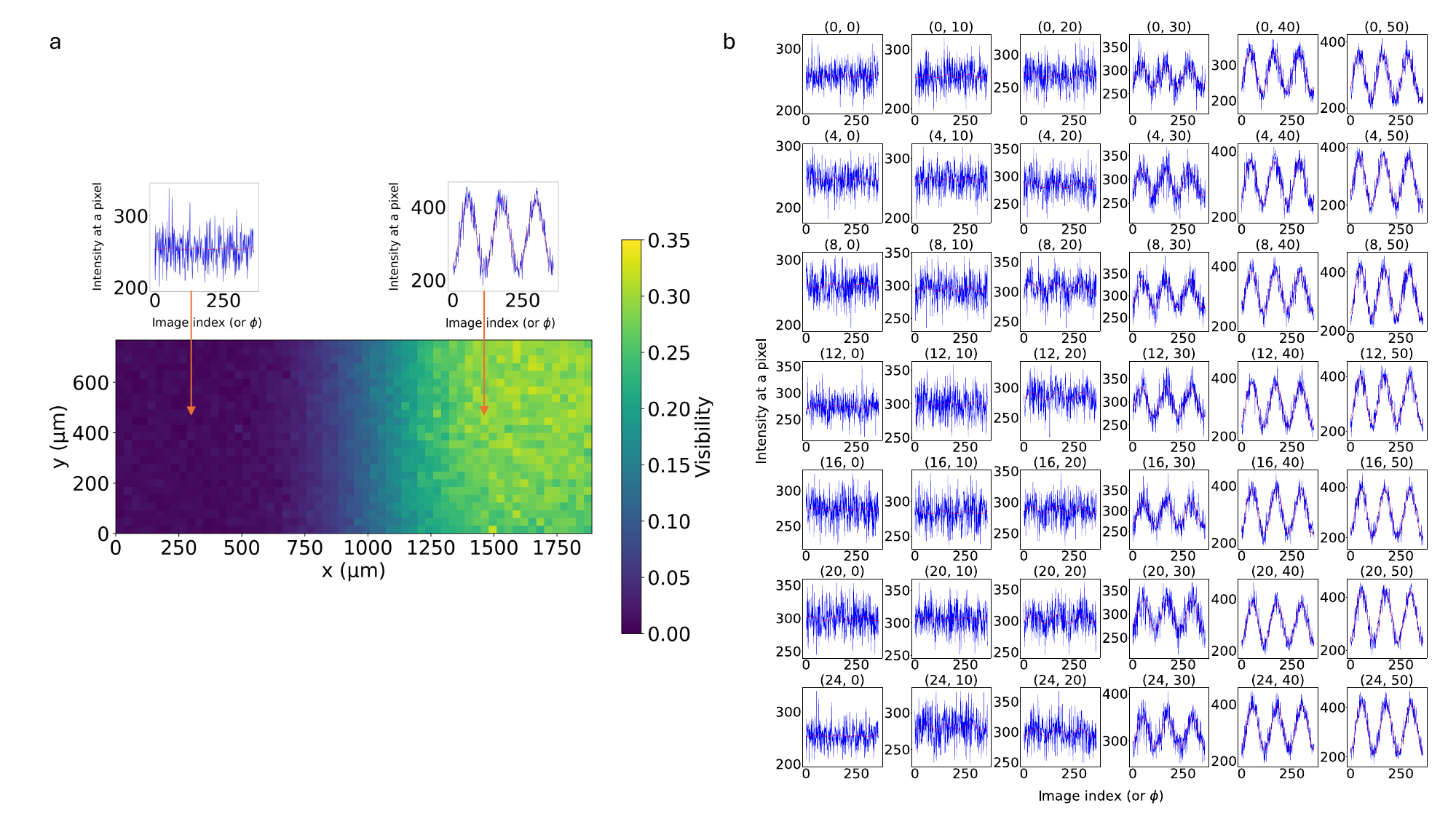}
		\caption{Experimental results for momentum space. (a) Represents
			the reconstructed visibility map of the knife edge -- an absorptive sharp object. (b) Illustration of the imaging process with the EMCCD camera capturing the detected beam intensity as the interferometric phase $\phi$ is scanned. The intensity varies with the phase change. By examining the recorded interference pattern on a pixel-by-pixel, independent information about each point on the object is derived. A sinusoidal function is fitted to the interference pattern of each pixel at various phases. This fitting process is demonstrated for a few pixels. The visibility for each pixel is determined from the obtained fit.}
		\label{fig:EPR-2d}
	\end{figure*}

    We place a knife-edge in front of the mirror $M_3$ in such a way that a part of the idler beam is blocked. An image of the knife-edge appears in the visibility of the interference pattern (Fig. ~\ref{fig:EPR-2d}a). This image can be interpreted as an edge-spread function. For the verification of position-momentum entanglement, we measure the corresponding edge-spread function in both momentum space (Fig.~\ref{fig_schematic}a) and position space (Fig.~\ref{fig_schematic}b), respectively (see Eqs. (\ref{photon rate final-mom}) and (\ref{pos-spread}) in Sec.~\ref{Sec:Theory} below and Fig.~\ref{fig:EPR-ESF_both_FF_and_NF} in Sec. ~\ref{Sec:Results} below). We access these two Fourier conjugate spaces (momentum and position correlations of signal and idler photons) with appropriate lens systems. 
\par
    In the momentum correlation measurement (Fig.~\ref{fig_schematic}a), we place a single lens of focal length $f= 200~\text{mm}$ in each arm of our interferometer. Here, the object (in our case, the knife-edge) and the camera are placed in the far field (Fourier plane) of the nonlinear crystal. The idler photons ($I_1$, $I_2$), generated during either the forward or backward pass are blocked and remain undetected by using a proper orientation of a polarizer placed in front of the EMCCD camera, permitting only signal photons ($S_1$, $S_2$) to pass through. 
\par    
    To measure the position correlation of the twin-photon pair (Fig.~\ref{fig_schematic}b), we use a 4f-imaging system, in every arm of our interferometer, consisting of a pair of lenses of focal lengths of $f = 100~\text{mm}$ such that both the knife-edge and the mirrors are in the near field (image plane) of the crystal with unit magnification \cite{Note-nf,tasca2009propagation}.
    
	\section{Theory}\label{Sec:Theory}
    We present a detailed theoretical analysis assuming that the two photons can have different wavelengths, i.e., the photon-generating process creates a non-degenerate energy spectrum. The degenerate case can be readily obtained by setting the wavelengths of the two photons equal in the resulting formulas. Throughout the analysis, we assume that photons propagate as paraxial beams and are incident normally on both the object and the detector. Our analysis is based on the theory developed through Refs.~\cite{lahiri2015theory,lahiri2017twin,viswanathan2021position}.   
    \par
    In the setup (Figs.~\ref{fig_schematic}a and \ref{fig_schematic}b), a single nonlinear crystal is weakly pumped from two opposite directions by two mutually coherent laser beams. The first pass of the pump (from left in Figs.~\ref{fig_schematic}) through the nonlinear crystal is equivalent to the source on the left in Fig.~\ref{fig:principle}b. Likewise, the second pass of the pump (from right in Figs.~\ref{fig_schematic}) through the nonlinear crystal is equivalent to the source on the right in Fig.~\ref{fig:principle}b. \emph{We denote the first and second passes of the pump by $Q_1$ and $Q_2$, respectively.}
    \par
    The twin-photon state generated by an individual pass ($j$) of the pump through nonlinear crystal can be expressed as
	\begin{equation} \label{state-spdc}
	|\psi_{j}\rangle = \int d\textbf{k}_{S}  \ d\textbf{k}_{I} \ C(\textbf{k}_{S}, \textbf{k}_{I}) |\textbf{k}_{S}\rangle_{S_{j}} |\textbf{k}_{I}\rangle_{I_{j}},
	\end{equation}
	where $j = 1,2$ ket $|\textbf{k}_{S}\rangle_{S_{j}}$ ($|\textbf{k}_{I}\rangle_{I_{j}}$) represents a signal (idler) photon with momentum $\hbar \textbf{k}_{s}$ ($\hbar \textbf{k}_{I}$) emitted by $Q_j$, and $C(\textbf{k}_{S}, \textbf{k}_{I})$ is the two-photon amplitude. The momentum correlation between the two photons is fully characterized by the joint probability density
	\begin{equation} \label{prob-mom}
	P(\textbf{k}_{S},\textbf{k}_{I}) = \left| C(\textbf{k}_{S}, \textbf{k}_{I}) \right|^{2},
	\end{equation}
 and the position correlation is fully characterized by the joint probability density 
	\begin{equation} \label{prob-pos}	P(\boldsymbol{\rho}_{S},\boldsymbol{\rho}_{I}) = \Big|\frac{1}{4\pi^{2}}\int d \textbf{k}_{S} \ d \textbf{k}_{I}\  C(\textbf{k}_{S}, \textbf{k}_{I}) \ e^{i (\textbf{k}_{S}.\boldsymbol{\rho}_{S} + \textbf{k}_{I}.\boldsymbol{\rho}_{I})}\Big|^{2}.
	\end{equation}
    These two probability densities are assumed to take the following Gaussian forms (cf. \cite{walborn2010spatial}; see also Appendix~\ref{app:A})
     \begin{subequations}
      \begin{align}
       P(\textbf{k}_{I},\textbf{k}_{S}) &\propto \text{exp} \left[-\frac{\sigma_{+}^{2}}{2} |\textbf{k}_{I} + \textbf{k}_{S}|^{2} \right], \label{joint-prob-approx-mom} \\
       P(\boldsymbol{\rho}_{S},\boldsymbol{\rho}_{I}) &\propto \text{exp}\left[ - \frac{2}{\sigma_{-}^{2}(1 + \lambda_{I}/\lambda_{S})^{2}}|\boldsymbol{\rho}_{S} - \boldsymbol{\rho}_{I}|^{2} \right], \label{joint-prob-pos-approx-nd}
      \end{align}   
     \end{subequations}
     where $\sigma_{+} = w_{p}$ with $w_{p}$ being the pump waist and $\sigma_{-} = \sqrt{L \lambda_{p} \lambda_{S}/(2 \pi\lambda_{I})}$ with $\lambda_{p}$, $\lambda_{I}$ and $\lambda_{S}$ being the wavelengths of the pump, idler and signal photons, respectively, and $L$ being the crystal length. 
     \par
     When the crystal is weakly pumped, the likelihood of multi-photon emission and stimulated emission is negligible compared to that of two-photon emissions \cite{wiseman2000induced,lahiri2019nonclassicality}. In this case, the quantum state of light created by the two sources is the linear superposition of two-photon states generated by each of them individually. It then follows from Eq.~(\ref{state-spdc}) that the resulting state is given by
    \begin{align} \label{state-spdc-super}
	&|\psi\rangle = \int d\textbf{k}_{S}  \ d\textbf{k}_{I} \ C(\textbf{k}_{S}, \textbf{k}_{I}) \cr
    &\times \left[ \hat{a}_{S_{1}}^{\dagger}(\textbf{k}_{S}) \hat{a}_{I_{1}}^{\dagger}(\textbf{k}_{I}) + e^{i\phi_p} \hat{a}_{S_{2}}^{\dagger}(\textbf{k}_{S}) \hat{a}_{I_{2}}^{\dagger}(\textbf{k}_{I})\right]|vac\rangle,
	\end{align}
    where the two passes of the pump through the crystal are labeled by $1$ and $2$, the quantity $\phi_p$ represents a phase, $\hat{a}_{S_j}^{\dagger}(\textbf{k}_S) \, \hat{a}_{I_j}^{\dagger}(\textbf{k}_{I})|vac\rangle =|\textbf{k}_{S}\rangle_{S_{j}} |\textbf{k}_{I}\rangle_{I_{j}}$ ($j=1,2$) with $\hat{a}^{\dag}$ and $|vac\rangle$ representing a photon creation operator and the vacuum state, respectively. We have also assumed for simplicity that the sources emit with the same probability and dropped the normalization factor $1/\sqrt{2}$ because it is irrelevant to our analysis. 
    \par
    As mentioned in Sec.~\ref{Sec:experiment}, the idler beam, $I_{1}$, emitted by $Q_{1}$, is aligned with the idler beam $I_{2}$ that is emitted by $Q_{2}$. An absorptive knife edge is placed in the idler beam between $Q_1$ and $Q_{2}$. The idler photon is not detected. The signal photon is detected after aligning the signal beams through the source. In order to determine $\Delta(\textbf{k}_{I}|\textbf{k}_{S})$ and $\Delta(\boldsymbol{\rho}_{I}|\boldsymbol{\rho}_{S})$, we obtain expressions for the interference patterns in the far-field and near-field \cite{Note-nf} configurations, respectively. In the former case, both the knife edge and the camera are placed on the Fourier plane of the sources (Fig.~\ref{fig_schematic}a) while in the latter case, they are placed in the image plane of the sources (Fig.~\ref{fig_schematic}b). 
	\par
    Suppose that $\boldsymbol{\rho}_{o}\equiv (x_o,y_o)$ represents a point on the object plane. An absorptive knife edge placed along the $x_o$-axis (i.e., $y_{o} = 0$) is represented by the amplitude transmission coefficient 
	\begin{equation} \label{knife edge def}
    T(\boldsymbol{\rho}_{o})=T(x_{o},y_o)=
	\begin{cases}
	0, & \text{if } x_{o} <0, \\
	1, & \text{if } x_{o} \geq 0.
	\end{cases}
	\end{equation} 
    When the knife edge is inserted in the idler beam between the two sources, the relationships between the idler fields generated by $Q_1$ and $Q_2$ are given by the following two equations for the far-field ($ff$) and near-field ($nf$) cases, respectively \cite{lahiri2015theory,viswanathan2021position}:
   \begin{subequations}
    \begin{align}
     &\hat{a}_{I_{2}}(\textbf{k}_{I}) = e^{i \phi_{I}}\left[F_{ff} T(\boldsymbol{\rho}_{o})\hat{a}_{I_{1}}(\textbf{k}_{I}) + R_{ff}(\boldsymbol{\rho}_{o}) \hat{a}_{0}(\textbf{k}_{I})\right], \label{alignment-mom} \\
      &\hat{E}_{I_{2}}^{(+)}(\boldsymbol{\rho}_{I}) = e^{i \phi_{I}}\big[F_{nf} T(\boldsymbol{\rho}_{o})\hat{E}_{I_{1}}^{(+)}(\boldsymbol{\rho}_{I}) \nonumber \\ & \qquad \qquad \qquad \qquad + R_{nf}(\boldsymbol{\rho}_{o}) \hat{E}_{0}^{(+)}(\boldsymbol{\rho}_{I})\big], \label{alignment-pos}
      \end{align}   
   \end{subequations}
where the phase $\phi_{I}$ is accumulated due to the propagation of the idler photon from $Q_{1}$ to $Q_{2}$. In Eq.~(\ref{alignment-mom}), a photon with transverse momentum $\hbar\textbf{k}_{I}$ impinges on the point $\boldsymbol{\rho}_{o}$, i.e., $\textbf{k}_{I} = (2 \pi /\lambda_{I} f) \boldsymbol{\rho}_{o}$ with $f$ being the focal length of the positive lens placed in idler path and $0\leq F_{ff} \leq 1$ characterizes losses of idler photons between the two sources due to misalignment and absorption/reflection at various optical components. When $F_{ff}=0$, the idler photon is lost with $100\%$ probability; and when $F_{ff}=1$, no idler photon is lost. Such losses and the interaction of the object with idler photons can be quantum mechanically treated like a beam splitter with single input \cite{zou1991induced}; in Eq. (\ref{alignment-mom}), $\hat{a}_{0}$ is the vacuum field at the unused port of the beam splitter (object) and the relation $R_{ff}(\boldsymbol{\rho}_{o})=\sqrt{1-[F_{ff}T(\boldsymbol{\rho}_{o})]^{2}}$ ensures that the photon number is conserved. Likewise, in the near-field case [Eq.~(\ref{alignment-pos})], we have $R_{nf}(\boldsymbol{\rho}_{o})=\sqrt{1-[F_{nf}T(\boldsymbol{\rho}_{o})]^{2}}$ and $\hat{E}_{0}^{(+)}(\boldsymbol{\rho}_{I})$ represents the vacuum field at the unused port of the beam splitter (object). In this case, $\hat{E}_{I_{j}}^{(+)}(\boldsymbol{\rho}_{I}) = \int d\textbf{k}_{I_{j}} \hat{a}_{I_{j}}(\textbf{k}_{I}) e^{i \textbf{k}_{I_{j}}\cdot \boldsymbol{\rho}_{I}}$ represents the propagating part of the idler field at source $Q_j$.
\par
Using Eqs.~(\ref{state-spdc-super}), (\ref{alignment-mom}), and (\ref{alignment-pos}), we obtain the following states for the far- and near-field cases: 
\begin{subequations}
    \begin{align}
     &|\psi\rangle_{ff} = \int d \textbf{k}_{S} \ d \textbf{k}_{I} \ C(\textbf{k}_{S}, \textbf{k}_{I}) \Big( |\textbf{k}_{S}\rangle_{S_{1}} \nonumber \\
     & \qquad \qquad + e^{i (\phi_{p}-\phi_{I})} F_{ff} T^{*}(\boldsymbol{\rho}_{o})| \textbf{k}_{S}\rangle_{S_{2}} \Big)| \textbf{k}_{I}\rangle_{I_{1}} \nonumber \\
     &+ \int d \textbf{k}_{S} \ d \textbf{k}_{I} \ C(\textbf{k}_{S}, \textbf{k}_{I}) e^{i (\phi_{p}-\phi_{I})} R^{*}_{ff} (\boldsymbol{\rho}_{o}) |\textbf{k}_{S}\rangle_{S_{2}}|\textbf{k}_{I}\rangle_{0}, \label{aligned state-mom} \\
      &|\psi \rangle_{nf} = \int d \textbf{k}_{I_{1}} \ d \textbf{k}_{S_{1}} C(\textbf{k}_{S}, \textbf{k}_{I}) e^{i (\phi_{p}-\phi_{I})} |\textbf{k}_{I_{1}}\rangle_{I_{1}} |\textbf{k}_{S_{1}}\rangle_{S_{1}} \cr
      &+ \int d \textbf{k}_{I_{2}} \ d \textbf{k}_{S_{2}} \ d \textbf{k}'_{I}\ C(\textbf{k}_{I_{2}}, \textbf{k}_{S_{2}}) \ e^{i (\phi_{p}-\phi_{I})} |\textbf{k}_{S_{2}}\rangle_{S_{2}} \cr
      &\otimes \big( F_{nf} \ \tilde{T}^{'*}(\textbf{k}_{I_{2}}- \textbf{k}'_{I}) \ |\textbf{k}'_{I}\rangle_{I_{1}} + \tilde{R}_{nf}^{'*}(\textbf{k}_{I_{2}}- \textbf{k}'_{I}) |\textbf{k}'_{I}\rangle_{0} \big), \label{aligned state-pos}
      \end{align}   
   \end{subequations}
where $\tilde{T}(\textbf{k}_{I})$ and $\tilde{R}(\textbf{k}_{I})$ are the Fourier transforms of $T(\boldsymbol{\rho}_{o})$ and $R(\boldsymbol{\rho}_{o})$, respectively; and we applied the convolution theorem to obtain Eq.~(\ref{aligned state-pos}) as shown in \cite{viswanathan2021position}.

\par
The signal photon is detected after aligning the signal beams through the source. Consequently, the positive frequency part of the quantized electric field, at a point $\boldsymbol{\rho}_{c} \equiv (x_{c},y_{c})$ on the camera plane, in the far-field and near-field configurations, respectively, can be represented by \cite{lahiri2015theory,lahiri2017twin,viswanathan2021position}
     \begin{subequations}
      \begin{align}
      \hat{E}^{(+)}_{ff}(\boldsymbol{\rho}_{c}) &\propto  \hat{a}_{S_1}(\textbf{k}_{s}) +e^{i \phi_{S}}\hat{a}_{S_2}(\textbf{k}_{S}), \label{detection-op-mom} \\
       \hat{E}_{nf}^{(+)}(\boldsymbol{\rho}_{c}) &\propto \int d\textbf{k}_{S} \left[ \hat{a}_{S_{1}}(\textbf{k}_{S}) +e^{i \phi_{S}} \hat{a}_{S_{2}}(\textbf{k}_{S}) \right] \ e^{i \textbf{k}_{S}.\boldsymbol{\rho}_{S}}, \label{detection-op-pos}
      \end{align}   
     \end{subequations}
     where $\phi_{S}$ is the phase accumulated due to the propagation of the signal photon from the crystal to the mirror (M2 in Fig.~\ref{fig_schematic}) and back to the crystal; in Eq.~(\ref{detection-op-mom}) $\textbf{k}_{S} = (2 \pi /\lambda_{S} f_{c}) \boldsymbol{\rho}_{c}$ with $f_{c}$ being the focal length of the positive lens placed in signal path; and in Eq.~(\ref{detection-op-pos}), $\boldsymbol{\rho}_{S}=\boldsymbol{\rho}_{c}/M_S$ with $M_S$ being the magnification of the imaging system that images the crystal on the camera along the signal path.
     \par
     We can now determine the single-photon counting rate (intensity), $\mathcal{R}(\boldsymbol{\rho}_{c})$, at a point $\boldsymbol{\rho}_{c}$ on the camera using the well-known formula introduced by Glauber \cite{glauber1963quantum}: $\mathcal{R}(\boldsymbol{\rho}_{c}) \propto \langle \psi|\hat{E}^{(-)}(\boldsymbol{\rho}_{c})\hat{E}^{(+)}(\boldsymbol{\rho}_{c})|\psi\rangle $, where $\hat{E}^{(-)}=\{ \hat{E}^{(+)} \}^{\dag}$. It follows from Eqs.~(\ref{aligned state-mom})--(\ref{detection-op-pos}), Eq.~(\ref{knife edge def}), and Eqs.~(\ref{prob-mom}) and (\ref{prob-pos}) that the single-photon counting rates in the far-field and near-field configurations are, respectively, given by (Appendix~\ref{app:B})
  \begin{subequations}
      \begin{align}
       &\mathcal{R}_{ff}(x_c,y_c) \nonumber \\ &  \propto 1 + \frac{1}{2} F_{ff} \left( 1- \text{Erf}\left[\frac{\sqrt{2}\pi \sigma_{+} }{f_{c} \lambda_{S}} x_c \right] \right) \cos \phi_{in}, \label{photon_count_rate_mom} \\
       &\mathcal{R}_{nf}(x_c,y_c) \nonumber \\ &  \propto 1 + \frac{1}{2} F_{nf}  \left(1+\text{Erf}\left[\frac{\sqrt{2}\ \lambda_{S}}{M_S(\lambda_{I}+\lambda_{S})\sigma_{-}} x_c \right] \right) \cos\phi_{in}, \label{photon_count_rate_pos_loss_absorb}
      \end{align}   
     \end{subequations}
    where $\phi_{in}= \phi_{p} - \phi_{I}- \phi_{S}$ is the tunable interferometric phase, $(x_c,y_c) \equiv \boldsymbol{\rho}_{c}$, and Erf represents the error function. It is evident that if $\phi_{in}$ is varied $\mathcal{R}_{ff}(\boldsymbol{\rho}_{c})$ and $\mathcal{R}_{nf}(\boldsymbol{\rho}_{c})$ vary sinusoidally. Therefore, Eqs.~\eqref{photon_count_rate_mom} and \eqref{photon_count_rate_pos_loss_absorb} represent interference patterns. We note that both $\mathcal{R}_{ff}$ and $\mathcal{R}_{nf}$ do not depend on $y_c$, which is expected because the knife edge was placed along $x_o$-axis on the object plane [see Eq.~\eqref{knife edge def}].
     \par
     If we now determine the visibility \cite{Note-vis} of the two interference patterns given by Eqs.~\eqref{photon_count_rate_mom} and \eqref{photon_count_rate_pos_loss_absorb}, we find that
     \begin{subequations}
      \begin{align}
       &\mathcal{V}_{ff}(x_c) =  \frac{F_{ff}}{2} \, \left( 1- \text{Erf}\left[\frac{\sqrt{2}\pi \sigma_{+} }{f_{c} \lambda_{S}} x_c \right] \right), \label{photon rate final-mom} \\
       &\mathcal{V}_{nf}(x_c) = \frac{F_{nf}}{2} \, \left( 1+\text{Erf}\left[\frac{\sqrt{2}\ \lambda_{S}}{M_s(\lambda_{I}+\lambda_{S})\sigma_{-}}x_c \right] \right). \label{pos-spread}
      \end{align}   
     \end{subequations}
     It can be noted that these visibilities give images of the knife edge in the far-field and near-field configurations; that is, each represents the edge-spread function (ESF) in the respective configuration.
     \par
     The visibilities (ESFs) given by Eqs.~(\ref{photon rate final-mom}) and (\ref{pos-spread}) can be used to measure the quantities $\sigma_{+}$ and $\sigma_{-}$. The spreads of these two ESFs ($D_k$ and $D_{\rho}$) can be defined by the inverse of the coefficients that multiply $x_c$ within the error function, i.e., 
     \begin{subequations}
      \begin{align}
       \sigma_{+}  &= \frac{f_{c} \lambda_{S}}{\sqrt{2} \pi} \,\frac{1}{D_{k}}, \label{sigmaplus-ESF-relation}\\
       \sigma_{-} &= \frac{\sqrt{2}\lambda_{S}}{M_S(\lambda_{I}+\lambda_{S})} D_{\rho}, \label{sigmaminus-ESF}
      \end{align}   
     \end{subequations} 
where we note again that $\sigma_{+} = w_{p}$ and $\sigma_{-} = \sqrt{L \lambda_{p} \lambda_{S}/(2 \pi\lambda_{I})}$. From the definition of the error function, it follows that these spreads ($D_k$ and $D_{\rho}$) are the distances for which the visibilities $\mathcal{V}_{ff}$ and $\mathcal{V}_{nf}$ rise from $24 \%$ to $76\%$ of their maximum values. Therefore, $D_k$ and $D_{\rho}$ are experimentally measurable quantities, which are related to $\sigma_{+}$ and $\sigma_{-}$ by Eqs.~\eqref{sigmaplus-ESF-relation} and \eqref{sigmaminus-ESF}, respectively. That is, $\sigma_{+}$ and $\sigma_{-}$ can be experimentally obtained from the visibility of the interference patterns. We now proceed to show how to verify the spatial (momentum-position) entanglement by testing both the EPR criterion and the MGVT criterion. 
\par
\emph{Testing the EPR criterion:} One can show from the first principle that the conditional variances in Eq.~\eqref{EPR def} are related to $\sigma_{+}$ and $\sigma_{-}$ by the following relations (Appendix~\ref{app:C})  \begin{subequations}\label{cond-var-mom-ps}
\begin{align} 
     &\Delta^{2}(k_{Ix}|k_{Sx}) \approx \frac{1}{\sigma_{+}^{2}}, \label{cond-var-mom-ps:a} \\
      &\Delta^{2}(x_{I}|x_{S}) \approx  \left[\frac{1+(\lambda_{I}/\lambda_{S})}{2} \right]^{2} \sigma_{-}^{2}. \label{cond-var-mom-ps:b}
\end{align}   
\end{subequations}
Using Eqs.~(\ref{sigmaplus-ESF-relation}), (\ref{sigmaminus-ESF}), (\ref{cond-var-mom-ps:a}) and (\ref{cond-var-mom-ps:b}), the product of the conditional variances in the EPR-criterion [Eq.~(\ref{EPR def})] can be expressed as
\begin{align} \label{Var-product-ESF}
       \Delta^{2}(x_{I}|x_{S}) \  \Delta^{2}(k_{Ix}|k_{Sx}) = \frac{\pi^{2}}{f_{c}^{2} \lambda_{S}^{2}M_{S}^{2}}D_{\rho}^{2}D_{k}^{2}.
\end{align}
In our method, the quantity on the right-hand side of Eq.~\eqref{Var-product-ESF} is determined from single-photon interference data. If this quantity is smaller than $1/4$, the photon pair is spatially entangled according to the EPR criterion. 
\par
\emph{Testing the MGVT criterion:} The quantities $\Delta^{2}K_{x+}$ and $ \Delta^{2}X_{-}$ appearing in the MGVT criterion [Eq.~(\ref{MGVT criteria-main})] can also be related to the experimentally measurable spreads of ESFs [see Appendix~\ref{app:D}, Eqs.~(\ref{variance-mom-MGVT}) and (\ref{variance-pos-MGVT})]. Using these relations, the product of the conditional variances in the MGVT criterion [Eq.~(\ref{MGVT criteria-main})] can be expressed as
\begin{align}\label{MGVT-ESF-main}
       \Delta^{2}K_{x+} \, \Delta^{2}X_{-} =\frac{\pi^{2}}{f_{c}^{2} \lambda_{S}^{2}M_{S}^{2}} \, D_{k}^{2} \ D_{\rho}^{2}.
\end{align}
Once again, the quantity on the right-hand side of Eq.~\eqref{MGVT-ESF-main} can be determined from single-photon interference data. If the value of this quantity is smaller than $1$, the photon pair is spatially entangled according to the MGVT criterion.
\par
We now briefly discuss the robustness of our method in the context of key experimental imperfections. The factors $F_{ff}$ and $F_{nf}$ \textemdash ~which characterize losses of idler photons and misalignment of idler beams in the far-field and near-field configurations, respectively \textemdash ~appear in the single-photon counting rates given by Eqs.~(\ref{photon_count_rate_mom}) and (\ref{photon_count_rate_pos_loss_absorb}). It is evident from these equations that the visibility of the interference patterns will depend on these factors: a higher loss in the idler path (i.e., smaller $F_{ff}$ and $F_{nf}$) results in a lower visibility of the signal interference patterns [Eqs.~(\ref{photon rate final-mom}) and (\ref{pos-spread})]. However, it is remarkable that the values of $\sigma_{+}$ and $\sigma_{-}$ obtained using these interference patterns do not depend on $F_{ff}$ and $F_{nf}$ as can be observed in Eqs.~(\ref{sigmaplus-ESF-relation}) and (\ref{sigmaminus-ESF}). Therefore, our method is applicable even when the interference-visibility is very low due to high loss of idler photons and/or idler beam misalignment. This fact is supported by the experimental results presented in the next section.
      
\section{Results} \label{Sec:Results}

	\begin{figure*}
		\centering 
		\includegraphics[width=1.0\linewidth]{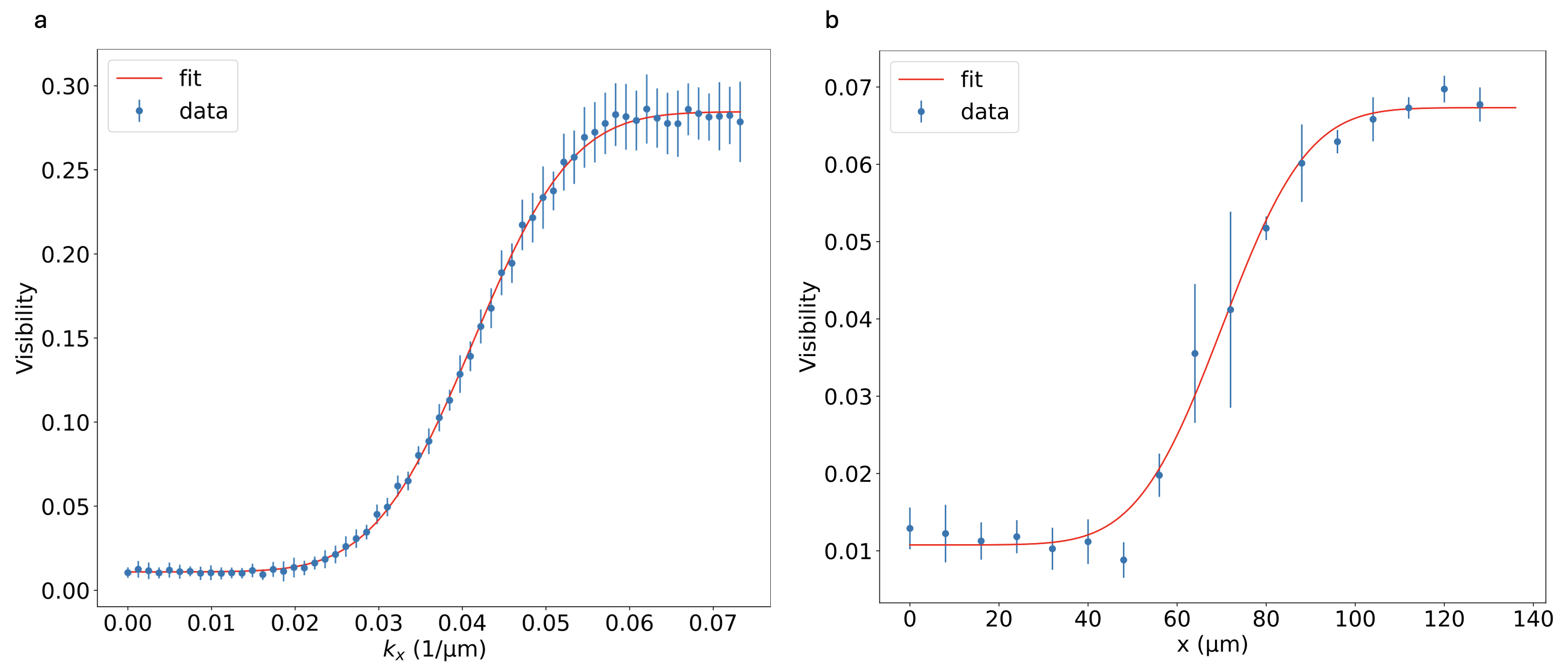}	
        \caption{
Quantitative measurement of the edge-spread function in momentum and position space. (a) Cross-section of the reconstructed visibility image obtained from a sharp knife-edge placed in the momentum plane. The visibility profile is fitted with an error function, and the characteristic width $D_k$ is extracted from the 76\%–24\% rise interval. The measured value is $D_k = (352 \pm 17)\,\mu\mathrm{m}$, in good agreement with the theoretical prediction of $337\,\mu\mathrm{m}$. 
(b) Corresponding measurement in the position plane. From the error-function fit, we obtain $D_\rho = (28 \pm 3)\,\mu\mathrm{m}$, compared with the theoretical value of $21\,\mu\mathrm{m}$. The reduced visibilities in both measurements reflect the significant losses present in the idler arm; nonetheless, the extracted widths $D_k$ and $D_\rho$ remain unaffected by these losses, demonstrating the robustness of the method.}
		\label{fig:EPR-ESF_both_FF_and_NF}
	\end{figure*}

In order to analyze the position-momentum entanglement of signal and idler photons generated at the nonlinear crystal, without relying on coincidence detection, we captured intensity variations of the detected beams while varying the interferometric phase $\phi_{in}$. We recorded a total of 360 frames, each associated with a value of $\phi_{in} = \phi_1, \phi_2, \ldots, \phi_{360}$. Subsequently, we analyzed the recorded interference patterns on a pixel-by-pixel basis to form a visibility image (see Fig.~\ref{fig:EPR-2d}). 
\par 
In the case of imperfect momentum correlation, a single point on the camera no longer corresponds to just one point on the object due to the varying momenta of undetected photons. As momentum correlation weakens, the range broadens resulting in information from multiple points on the object appearing at a single point on the camera. This indicates that a weaker momentum correlation decreases the spatial resolution. 
\par 
In order to evaluate the variance in the far field, we use the visibility image of the knife edge. From the cross-section of the position-dependent visibility image, we average over 20 pixel-rows (Fig.~\ref{fig:EPR-2d}a) and fit it with the error function. The measured visibility with respect to distance in the momentum space is shown in Fig.~\ref{fig:EPR-ESF_both_FF_and_NF}a. Quantitatively, the spread of the ESF is given by the parameter $D_k$ which is the characteristic distance, on the camera plane, over which the visibility rises from $24 \%$ to $76\%$ of its maximum value (Fig. \ref{fig:EPR-ESF_both_FF_and_NF}a). This experimentally measured value of $D_k$ is $352 \pm 17$ $\mu$m, which is in good agreement with the corresponding theoretical value $D_k = 337 \mu m$. The theoretical value is obtained from Eq.~(\ref{sigmaplus-ESF-relation}) by substituting the following values for the experimental parameters: $\sigma_+=w_{p} = 108\, \mu m$, $\lambda_{S}= 810\,nm$, and $f_c=200\, mm$. Note that our method of measuring $D_k$, is an alternative to the method some of us introduced in Refs.~\cite{lahiri2017twin, hochrainer2017quantifying}, which was limited to the measurement of momentum correlations.
\par
Similarly, to determine the variance in the near field (position correlation), the intensity data for each pixel was analyzed and the visibility was calculated. The visibility image of the knife edge is depicted in Fig.~\ref{fig:EPR-ESF_both_FF_and_NF}b. For position measurements, both the knife edge and the camera are on the image plane of the crystal. Based on the error function fit in Fig.~\ref{fig:EPR-ESF_both_FF_and_NF}b, $D_{\rho}$ which is the $24\%-76\%$ width of the edge spread function, turns out to be $28\pm3$ $\mu$m. The theoretical value of $D_{\rho}$ obtained using Eq.~(\ref{sigmaminus-ESF}) is $16$ $\mu$m; here $\sigma_{-} = \sqrt{L \lambda_{p} \lambda_{S}/(2 \pi\lambda_{I})}$ and the experimental parameters are $L=2\,mm$, $\lambda_{p}=405\, nm$, $\lambda_{S} = \lambda_{I} = 810\, nm$, and $M_s=1$. This shows that the experimental value matches the theoretical value with good accuracy. 
\par
From the experimental results, i.e., measurements of  $D_\rho$ and $D_k$ we calculate, using Eq. (\ref{Var-product-ESF}), the product of the conditional variances in  momentum and position to be
\begin{align} \label{EPR-exp-res}
       \Delta^{2}(x_{I}|x_{S}) \  \Delta^{2}(k_{Ix}|k_{Sx}) &= (3.30 \pm 0.75 )\times 10^{-2} \nonumber \\ &< \frac{1}{4},  
      \end{align}
which shows a violation of the EPR criterion [Eq. (\ref{EPR def})]. The corresponding theoretical value of the product of the conditional variances is $0.011<1/4$. This violation certifies the position-momentum entanglement of the signal and idler photons produced in the SPDC process. 
\par	
To strengthen our results further, we have also verified the violation of the MGVT separability criterion. The theoretical result is obtained by using Eqs.~(\ref{variance-mom-MGVT}) and (\ref{variance-pos-MGVT}) of Appendix~\ref{app:D}, and is found to be $\Delta^{2}K_{x+} \, \Delta^{2}X_{-}=0.011 <1$. The experimental value of the product is obtained using the experimentally measured values of $D_{k}$ and $D_{\rho}$. It is given by $\Delta^{2}K_{x+} \, \Delta^{2}X_{-} = (3.30 \pm 0.75 )\times 10^{-2} < 1$, 
which demonstrates the violation of the MGVT criterion.
\par
The measured product of conditional variances in both position and momentum spaces are in good agreement with the corresponding theoretical values. The small difference is due to experimental limitations, especially due to imperfections in the distances used in the imaging system, non-negligible lens thickness, imperfect positioning of knife edge, and relatively large camera pixel size. These affect more prominently the results for near-field correlations. 

\section{Discussion}\label{Sec:discuss}
Our method is highly resistant to key experimental imperfections. Dominating imperfections in our experiment are loss of idler photons at various optical components and misalignment of idler beams. These imperfections reduce the visibility of single-photon interference patterns and cannot be compensated for in any way. We have provided a theoretical justification at the end of Sec.~\ref{Sec:Theory} explaining why our method is resistant to such experimental imperfections. Our experimental data confirm significant presence of these imperfections during the experiment: the maximum value of the visibility in the far-field configuration is less than $0.3$ (Fig.~\ref{fig:EPR-ESF_both_FF_and_NF}a) and that in the near-field configuration is less than $0.07$ (Fig.~\ref{fig:EPR-ESF_both_FF_and_NF}b). Despite the presence of such high level of experimental imperfections, our method has worked remarkably well. 
\par
 Our technique is fundamentally different from the one used in \cite{bhattacharjee2022measurement} where EPR correlations in a bi-photon system have been studied with only single-photon detection. Unlike \cite{bhattacharjee2022measurement},  we do not measure the cross-spectral density, nor do we construct the conditional probability densities to measure the variances. Instead, we determine the variances directly from the visibility data.  
 \par
 We consider a two-photon pure state in our investigation to certify spatial entanglement in the system. However, it should be possible to extend the analysis to verify mixed state continuous variable entanglement using our method. This interferometric technique with undetected photons has been used to characterize mixed state entanglement in discrete variables (i.e., polarization degree of freedom) \cite{lahiri2021characterizing,lemos2023one,rajeev2023single,rajeev2025complete}. 

\section{Summary}\label{Sec:Summary}

 We have demonstrated both theoretically and experimentally that using an  interferometric technique, it is possible to verify entanglement in an infinite-dimensional bipartite photonic system simply without any coincidence measurement or postselection. From the visibility data obtained from single-photon interference patterns, we have been able to study the EPR correlations and demonstrate the violation of the MGVT criterion. Our investigation yields good agreement between the theoretical values and the experimentally measured results. Furthermore, since our method does not involve any coincidence detection, it is naturally well suited to study entanglement in highly non-degenerate sources, particularly when suitable detectors are not available for one of the wavelengths. 
 
\section*{Author contribution}
M.L. conceived the idea of the project, supervised the theoretical part of the project, and discussed the results. S.K. built and implemented the experimental setup, performed all measurements, developed the data analysis software, and carried out the full experimental data analysis. B.V. carried out the theoretical analysis and discussed the results. R.L. designed the setup, discussed the results and supervised the experimental part of the project. G.B.L. discussed the implementation, results,  and advised S.K. throughout the entire experimental implementation, including the setup, the alignment and the data collection. P.S. reviewed the data analysis and discussed the results. B.V., S.K. and M.L. wrote the manuscript with contributions of all the authors. 
 
\begin{acknowledgments}
We thank Inna Kviatkovsky for fruitful discussions about the quantum imaging setup. This material is based upon work supported by the Air Force Office of Scientific Research under award number FA9550-23-1-0216. S.K., P.S., and R.L. acknowledge the support of the following funding agencies: Foundation for Polish Science (FIRST TEAM project FENG.02.02-IP.05-0253/23); Narodowe Centrum Nauki (grant 2022/47/B/ST7/03465); National Centre for Research and Development QuantERA II projects: QM3 (QuantERAII/02/QM3/03/2024) and EXTRASENS (QuantERAII/02/EXTRASENS/02/2024); HORIZON EUROPE Marie Skłodowska-Curie Actions (FLORIN ID 101086142).
\end{acknowledgments}
	
\appendix

\section{Probability distributions corresponding to twin-photon momentum and position correlations generated by SPDC}\label{app:A}
The joint probability distribution of the twin photons, produced in degenerate SPDC \cite{tasca2009propagation, schneeloch2016introduction}, with the double Gaussian approximation, can be generalized to the non-degenerate case. In the momentum space, the joint probability density is written as
\begin{align} \label{jointprob-mom-nd}
	P(\textbf{k}_{S}, \textbf{k}_{I}) &= \frac{\sigma_{+}^{2} \sigma_{-}^{2}}{4 \pi^{2} } \left(1 + \frac{\lambda_{I}}{\lambda_{S}}\right)^{2} \text{exp}\left[-\frac{\sigma_{+}^{2}}{2 }|\textbf{k}_{I} + \textbf{k}_{S}|^{2}\right] \cr
	&\times \text{exp}\left[-\frac{\sigma_{-}^{2}}{2 }\left|\textbf{k}_{S} -\frac{\lambda_{I}}{\lambda_{S}} \textbf{k}_{I}\right|^{2}\right],
\end{align}
where $\sigma_{+} = w_{p}$ with $w_{p}$ being the pump waist and $\sigma_{-} = \sqrt{L \lambda_{p} \lambda_{S}/(2 \pi\lambda_{I})}$ with $\lambda_{p}$, $\lambda_{I}$ and $\lambda_{S}$ being the wavelengths of the pump, idler and signal photons, respectively, and $L$ being the crystal length. For the typical values of parameters in experiments, the second exponent in Eq. (\ref{jointprob-mom-nd}) varies very slowly compared to the first exponent. It is customary to treat the second exponent as a constant. Therefore, in the far-field scenario, we can reduce Eq.~(\ref{jointprob-mom-nd}) to
\begin{equation} \label{jointprob-mom-nd-approx}
P(\textbf{k}_{S}, \textbf{k}_{I}) \propto  \text{exp}\left[-\frac{\sigma_{+}^{2}}{2 }|\textbf{k}_{I} + \textbf{k}_{S}|^{2}\right],  
\end{equation}
which is Eq.~\eqref{joint-prob-approx-mom} in the main text. 
\par
The joint probability distribution of the twin photons in the position space is written as
\begin{align} \label{jointprob-pos-nd}
&P(\boldsymbol{\rho}_{S}, \boldsymbol{\rho}_{I}) = \left(\frac{4}{\pi^{2} \sigma_{+}^{2}\sigma_{-}^{2}(1 +\lambda_{I}/\lambda_{S})^{2}}\right) \cr
&\times \text{exp}\left[ - \frac{2}{\sigma_{+}^{2}(1 + \lambda_{I}/\lambda_{S})^{2}}|\boldsymbol{\rho}_{I} +(\lambda_{I}/\lambda_{S}) \boldsymbol{\rho}_{S}|^{2} \right] \cr
&\times \text{exp}\left[ - \frac{2}{\sigma_{-}^{2}(1 + \lambda_{I}/\lambda_{S})^{2}}|\boldsymbol{\rho}_{S} - \boldsymbol{\rho}_{I}|^{2} \right].
\end{align}
For typical experimental parameters, the first exponent in Eq. (\ref{jointprob-pos-nd}) varies very slowly compared to the second one. Because of this, we can treat the first exponent as a constant and approximate Eq.~(\ref{jointprob-pos-nd}) by
\begin{equation} \label{approx-joint-prob-pos}
P(\boldsymbol{\rho}_{S}, \boldsymbol{\rho}_{I}) \propto \text{exp}\left[ - \frac{2}{\sigma_{-}^{2}(1 + \lambda_{I}/\lambda_{S})^{2}}|\boldsymbol{\rho}_{S} - \boldsymbol{\rho}_{I}|^{2} \right],  
\end{equation}
which is Eq.~\eqref{joint-prob-pos-approx-nd} in the main text.

\section{Derivations of Eqs.~\eqref{photon_count_rate_mom} and \eqref{photon_count_rate_pos_loss_absorb}}\label{app:B}
We first present the derivation of Eq.~\eqref{photon_count_rate_mom}, which gives an expression of the single-photon counting rate at the camera obtained in the far-field case. In this case, the single-photon counting rate is given by \cite{glauber1963quantum} $\mathcal{R}_{ff}(\boldsymbol{\rho}_{c}) \propto \, _{ff}\langle \psi|\hat{E}_{ff}^{(-)}(\boldsymbol{\rho}_{c})\hat{E}_{ff}^{(+)}(\boldsymbol{\rho}_{c})|\psi\rangle_{ff} $, where $\hat{E}_{ff}^{(-)}=\{ \hat{E}_{ff}^{(+)} \}^{\dag}$. Substituting for $|\psi\rangle_{ff}$ and $\hat{E}_{ff}^{(+)}$ from Eqs.~\eqref{aligned state-mom} and \eqref{detection-op-mom}, respectively, we find that
\begin{align} \label{Photon_Counting Rate_Def_FF}
\mathcal{R}_{ff}(\boldsymbol{\rho}_{c}) \propto  \int d \textbf{k}_{I} \, P(\textbf{k}_{S},\textbf{k}_{I}) \left(1+ F_{ff} \, |T(\boldsymbol{\rho}_{o})|   \ \cos \phi_{in}\right).
\end{align}
Now substituting for $T(\boldsymbol{\rho}_{o})$ and $P(\textbf{k}_{S},\textbf{k}_{I})$ from Eqs.~(\ref{knife edge def}) and \eqref{jointprob-mom-nd-approx} into Eq.~\eqref{Photon_Counting Rate_Def_FF}, we obtain Eq.~\eqref{photon_count_rate_mom} in the main text:
\begin{align} \label{R-final-FF}
&\mathcal{R}_{ff}(x_c,y_c) \nonumber \\ &  \propto 1 + \frac{1}{2} F_{ff} \left( 1- \text{Erf}\left[\frac{\sqrt{2}\pi \sigma_{+} }{f_{c} \lambda_{S}} x_c \right] \right) \cos \phi_{in},
\end{align}
where $(x_{c},y_{c})\equiv \boldsymbol{\rho}_{c}$. 
\par
We now show the derivation of Eq.~\eqref{photon_count_rate_pos_loss_absorb}, which gives an expression for the single-photon counting rate at the camera obtained in the near-field case. In this case, the single-photon counting rate is given by $\mathcal{R}_{nf}(\boldsymbol{\rho}_{c}) \propto \, _{nf}\langle \psi|\hat{E}_{nf}^{(-)}(\boldsymbol{\rho}_{c})\hat{E}_{nf}^{(+)}(\boldsymbol{\rho}_{c})|\psi\rangle_{nf} $, where $\hat{E}_{nf}^{(-)}=\{ \hat{E}_{nf}^{(+)} \}^{\dag}$. Substituting for $|\psi\rangle_{nf}$ and $\hat{E}_{nf}^{(+)}$ from Eqs.~\eqref{aligned state-pos} and \eqref{detection-op-pos}, respectively, we find that
\begin{align} \label{Photon_Counting Rate_Def_NF}
\mathcal{R}_{nf}(\boldsymbol{\rho}_{c}) \propto \int d \boldsymbol{\rho}_{o} P(\boldsymbol{\rho}_{c},\boldsymbol{\rho}_{o}) \ \left(1+F_{nf} \, |T(\boldsymbol{\rho}_{o})| \cos \phi_{in} \right),
\end{align}
where $\boldsymbol{\rho}_{c}=M_S \boldsymbol{\rho}_{S}$ and $\boldsymbol{\rho}_{o}=M_I \boldsymbol{\rho}_{I}$ with $M_S$ being the magnification of the imaging system that images the crystal on the camera along the signal path and $M_I$ being the magnification of the imaging system that images the crystal on the knife edge along the idler path. Now using expressions for $P(\boldsymbol{\rho}_{c},\boldsymbol{\rho}_{o})$ [Eq.~\eqref{approx-joint-prob-pos}] and $T(\boldsymbol{\rho}_{o})$ [Eq.~(\ref{knife edge def})] we obtain the following relation from Eq.~(\ref{Photon_Counting Rate_Def_NF}):
\begin{align} \label{R-final-NF}
&\mathcal{R}_{nf}(x_c,y_c) \nonumber \\ &  \propto 1 + \frac{1}{2} F_{nf}  \left(1+\text{Erf}\left[\frac{\sqrt{2}\ \lambda_{S}}{M_S(\lambda_{I}+\lambda_{S})\sigma_{-}} x_c \right] \right) \cos\phi_{in}, 
\end{align}
which is Eq.~\eqref{photon_count_rate_pos_loss_absorb} in the main text.
 
\section{Derivations of Eqs.~\eqref{cond-var-mom-ps:a} and \eqref{cond-var-mom-ps:b}}\label{app:C}
For a two-photon system, the conditional variance in both momentum and position is given by \cite{howell2004realization} 
	\begin{subequations}
		\begin{align}
		\Delta^{2}(k_{Ij}|k_{Sj}) &= \int k_{Ij}^{2} \ P(k_{Ij}|k_{Sj}) \ d k_{Ij}\cr
		&- \Big(\int k_{Ij} \ P(k_{Ij}|k_{Sj}) \ d k_{Ij} \Big)^{2}, \label{variance-mom} \\
		\Delta^{2}(\rho_{Ij}|\rho_{Sj}) &= \int \rho_{Ij}^{2} \ P(\rho_{Ij}|\rho_{Sj}) \ d \rho_{Ij}\cr
		&- \Big(\int \rho_{Ij} \ P(\rho_{Ij}|\rho_{Sj}) \ d \rho_{Ij} \Big)^{2}, \label{variance-pos}
		\end{align}
	\end{subequations}
where $j=x,y$.
	\par
     The conditional probability density, $P(\textbf{k}_{I}|\textbf{k}_{S})$, can be obtained from the joint probability distribution Eq.~(\ref{jointprob-mom-nd}) using the standard formula: $P(\textbf{k}_{I}|\textbf{k}_{S}) = P(\textbf{k}_{I},\textbf{k}_{S})/ P(\textbf{k}_{S})$ which gives 
	\begin{align} \label{condprob-mom-nd}
	P(\textbf{k}_{I}|\textbf{k}_{S}) &= \frac{1}{2 \pi} \left[\sigma_{+}^{2} +  \left(\frac{\lambda_{I}}{\lambda_{S}} \right)^{2} \sigma_{-}^{2}\right] \cr
	&\times \text{exp}\left[-\frac{1}{2} \frac{(\sigma_{+}^{2} - (\lambda_{I}/\lambda_{S}) \sigma_{-}^{2})^{2}}{\sigma_{+}^{2} + (\lambda_{I}/\lambda_{S})^{2} \sigma_{-}^{2}}|\textbf{k}_{S}|^{2}\right] \cr &\times \text{exp}\left[ -\frac{1}{2} (\sigma_{+}^{2} + (\lambda_{I}/\lambda_{S})^{2}\sigma_{-}^{2}) |\textbf{k}_{I}|^{2}\right] \cr &\times \text{exp}\left[-(\sigma_{+}^{2} - (\lambda_{I}/\lambda_{S}) \sigma_{-}^{2})(\textbf{k}_{I}. \textbf{k}_{S}) \right].
	\end{align}
	The conditional variance in momentum is formally determined using Eqs.~(\ref{variance-mom}) and (\ref{condprob-mom-nd}). It is found to be given by 
	\begin{align} \label{cond-var-mom-nd}
	\Delta^{2}(k_{I_{j}}|k_{S_{j}}) &= \frac{1}{\sigma_{+}^{2} + (\lambda_{I}/\lambda_{S})^{2} \sigma_{-}^{2}}.     
	\end{align}
    For typical values of the parameters in experiments, it turns out that  $\sigma_{+}^{2} + (\lambda_{I}/\lambda_{S})^{2}\sigma_{-}^{2} \approx \sigma_{+}^{2} $. Thus, Eq. (\ref{cond-var-mom-nd}) can be expressed in the approximated form
    \begin{align} \label{cond-var-mom-nd-simp}
     \Delta^{2}(k_{I_{j}}|k_{S_{j}}) & \approx \frac{1}{\sigma_{+}^{2}}.       
    \end{align} 
By setting $j=x$ in Eq.~\eqref{cond-var-mom-nd-simp}, we obtain Eq.~\eqref{cond-var-mom-ps:a} in the main text.
\par
We now derive Eq.~\eqref{cond-var-mom-ps:b}. The conditional probability density, $P(\boldsymbol{\rho}_{I}|\boldsymbol{\rho}_{S})$, can be obtained from the joint probability distribution Eq.~(\ref{jointprob-pos-nd}) using the standard formula: $P(\boldsymbol{\rho}_{I}|\boldsymbol{\rho}_{S}) = P(\boldsymbol{\rho}_{I},\boldsymbol{\rho}_{S})/ P(\boldsymbol{\rho}_{S})$ which gives 
	\begin{align} \label{condprob-pos-nd}
	&P(\boldsymbol{\rho}_{I}|\boldsymbol{\rho}_{S}) = \frac{2}{\pi} \left(\frac{1}{1 +(\lambda_{I}/\lambda_{S})} \right)^{2} \left(\frac{\sigma_{+}^{2} + \sigma_{-}^{2}}{\sigma_{+}^{2} \sigma_{-}^{2}} \right) \cr
	&\times \text{exp}\left[-\frac{2}{(1 + \lambda_{I}/\lambda_{S})^{2}} \frac{(\sigma_{+}^{2} -(\lambda_{I}/\lambda_{S}) \sigma_{-}^{2})^{2}}{\sigma_{+}^{2}\sigma_{-}^{2}(\sigma_{+}^{2} + \sigma_{-}^{2})}|\boldsymbol{\rho}_{S}|^{2}\right] \cr
	&\times \text{exp} \left[- \frac{2}{(1 + \lambda_{I}/\lambda_{S})^{2}} \left(\frac{1}{\sigma_{+}^{2}} + \frac{1}{\sigma_{-}^{2}}\right)|\boldsymbol{\rho}_{I}|^{2}\right] \cr &\times \text{exp} \left[- \frac{4}{(1 + \lambda_{I}/\lambda_{S})^{2}} \right. \cr
    &\left. \qquad \quad\times \left(\frac{(\lambda_{I}/\lambda_{S})}{\sigma_{+}^{2}} + \frac{1}{\sigma_{-}^{2}}\right)(\boldsymbol{\rho}_{I}.\boldsymbol{\rho}_{S})\right].
	\end{align}
	The conditional variance in the position space is obtained using Eqs. (\ref{variance-pos}) and (\ref{condprob-pos-nd}). It is given by
	\begin{align} \label{cond-var-pos-nd}
	\Delta^{2}(\rho_{I_{j}}|\rho_{S_{j}}) &=  \left[\frac{1+(\lambda_{I}/\lambda_{S})}{2} \right]^{2}\left(\frac{\sigma_{+}^{2} \sigma_{-}^{2}}{\sigma_{+}^{2} + \sigma_{-}^{2}}\right). 
	\end{align}
	Using the same approximation that we used to simplify the conditional variance in momentum [see the discussion below Eq. (\ref{cond-var-mom-nd})], we reduce Eq. (\ref{cond-var-pos-nd}) to
    \begin{align} \label{cond-var-pos-nd-simp-0}
     \Delta^{2}(\rho_{I_{j}}|\rho_{S_{j}}) \approx \left[\frac{1+(\lambda_{I}/\lambda_{S})}{2} \right]^{2} \sigma_{-}^{2}
    \end{align}
    By setting $j=x$ in Eq.~\eqref{cond-var-pos-nd-simp-0}, we obtain Eq.~\eqref{cond-var-mom-ps:b} of the main text.

\section{Derivation of Eq.~\eqref{MGVT-ESF-main}}\label{app:D}
In our case, the MGVT criterion (Eq.~\eqref{MGVT criteria-main} in the main text) takes the form 
\begin{equation} \label{MGVT-app}
	\Delta^{2}K_{x+} \, \Delta^{2}X_{-} \geq 1,
\end{equation}
where $K_{x+} = K_{Ix} + K_{Sx}$, $X_{-} = x_{S} - x_{I}$, and 
\begin{subequations}
\begin{align}
	\Delta^{2}K_{x+} &= \int (K_{x+})^{2} \ P(k_{Sx},k_{Ix}) \ d k_{Ix}\cr
	&- \Big(\int K_{x+} \ P(k_{Sx},k_{Ix}) \ d k_{Ix} \Big)^{2}, \label{variance-K} \\
	\Delta^{2}X_{-} &= \int (X_{-})^{2} \ P(x_{S},x_{I}) \ d x_{I}\cr
	&- \Big(\int X_{-} \ P(x_{S},x_{I}) \ d x_{I} \Big)^{2}, \label{variance-X}
\end{align}
\end{subequations}
with $P(k_{Ix},k_{Sx})$ and $P(x_S,x_I)$ by given by Eqs.~(\ref{jointprob-mom-nd}) and \eqref{jointprob-pos-nd}, respectively. Carrying out these integrations, we obtain
\begin{subequations}
\begin{align}
&\Delta^{2}K_{x+}=\frac{2 \pi^2}{f_{c}^{2}\lambda_{S}^{2}} D_{k}^{2} \label{variance-mom-MGVT} \\
&\Delta^{2}X_{-}= \frac{1}{2 M_{S}^{2}}D_{\rho}^{2}, \label{variance-pos-MGVT}
\end{align}
\end{subequations}
where we applied Eqs.~\eqref{sigmaplus-ESF-relation} and \eqref{sigmaminus-ESF} from the main text and used the relations $\sigma_{+} = w_{p}$ and $\sigma_{-} = \sqrt{L \lambda_{p} \lambda_{S}/(2 \pi\lambda_{I})}$.
\par
Using Eqs.~\eqref{variance-mom-MGVT} and \eqref{variance-pos-MGVT}, we find that the left-hand side of Eq.~\eqref{MGVT-app} is given by
\begin{equation} \label{MGVT-ESF}
\Delta^{2}K_{+} \, \Delta^{2}X_{-} =\frac{\pi^{2}}{f_{c}^{2} \lambda_{S}^{2}M_{S}^{2}} \ D_{k}^{2} \ D_{\rho}^{2},   
\end{equation}
which is Eq.~\eqref{MGVT-ESF-main} in the main text.

\bibliography{EPR-bib}
	
\end{document}